\documentclass[11pt]{article}%
\usepackage{amssymb,amsmath,amsfonts,amsthm,array,bm,bbm,color,stmaryrd}%
\usepackage{amsmath}%
\usepackage{amsfonts}%
\usepackage{amssymb}%
\usepackage{graphicx}
\providecommand{\U}[1]{\protect\rule{.1in}{.1in}}

\numberwithin{equation}{section}

\newtheorem{theorem}{Theorem}[section]

\newtheorem{remark}[theorem]{Remark}

\def\<{\langle}
\def\>{\rangle}

\begin{document}

\title{Eddy heat exchange at the boundary under white noise turbulence}

\author{Franco Flandoli\footnote{Email: franco.flandoli@sns.it. Scuola Normale Superiore of Pisa, Piazza dei Cavalieri 7, 56124 Pisa, Italy.}
\quad Lucio Galeati\footnote{Email: lucio.galeati@iam.uni-bonn.de. Institute for Applied Mathematics, University of Bonn, Endenicher Allee 60, 53115 Bonn, Germany.}
\quad Dejun Luo\footnote{Email: luodj@amss.ac.cn. Key Laboratory of RCSDS, Academy of Mathematics and Systems Science, Chinese Academy of Sciences, Beijing 100190, China, and School of Mathematical Sciences, University of the Chinese Academy of Sciences, Beijing 100049, China. } }

\maketitle

\begin{abstract}
We prove the existence of an eddy heat diffusion coefficient coming from an
idealized model of turbulent fluid. A difficulty lies in the presence of a
boundary, with also turbulent mixing and the eddy diffusion coefficient going
to zero at the boundary. Nevertheless enhanced diffusion takes place.
\end{abstract}



\section{Introduction}

Eddy viscosity and eddy diffusion are two recognized phenomena which appear in
experiments and real situations under suitable fluid regimes. In this note we
focus on a particular case, but of main interest: the fact that the heat
exchange through a boundary may be increased by turbulence of the conducting
fluid. The problem can be investigated using different models from the one
used here, see e.g. Boussinesq problem. Here we model the phenomenon in the
following very simplified way: temperature $T=T\left(  t,\mathbf{x}\right)  $
is subject to the equation%
\begin{align}
\partial_{t}T &  =\kappa\Delta T+\mathbf{u}_{Q}\circ\mathbf{\nabla}%
T \qquad \text{in }\left[  0,T\right]  \times D\label{SPDE}\\
T|_{\partial D} &  =0,\qquad T|_{t=0}=T_{0}\qquad \text{in }D\nonumber
\end{align}
in a open connected bounded domain $D\subset\mathbb{R}^{d}$ with piecewise
regular boundary; $\kappa>0$ is the diffusion constant, that we should think
to be small; the velocity field, similarly to investigations for passive
scalars \cite{CertFal, FrischMazzino, GawKup, Kr, MajdaK, Sreen}, is a
given random, divergence free, vector field $\mathbf{u}_{Q}(
t,\mathbf{x})  $, Gaussian, white noise in time, with a prescribed covariance
matrix function $Q(  \mathbf{x},\mathbf{y} )  $ in space,
simulating in a simplified fashion an incompressible turbulent fluid. The
property of being white in time is certainly artificial compared to real
fluids;\ we consider this investigation a first step, to be completed in the
future with the understanding of more realistic regimes. We aim to recognize
in a quantitative way that, due to the random turbulent transport, heat
diffusion is enhanced.

Without noise and fluid motion, the temperature would decay to zero due to the
Dirichlet boundary conditions (the cold boundary absorbs heat) but the rate of
decay would be given by $\kappa\lambda_{D}$, where $-\lambda_{D}$ is the first
eigenvalue of the Laplacian operator $\Delta$ with zero boundary condition. But when the
fluid is turbulent, we expect a faster decay.

There are specific technical difficulties due to the boundary that we have to
overcome to prove the result. One problem is that the fluid fluctuations are
at rest on $\partial D$ (e.g. \cite{Liu-Pletcher, Wang-Pan-Wang})
namely $Q\left(  \mathbf{x,x}\right)  =0$ for $\mathbf{x}\in\partial D$. Hence
the strength of the mixing mechanism is depleted near the boundary, exactly
where the fluid comes in interaction with the cold boundary which is
responsible for cooling. We therefore have to understand the balance between
these phenomena.

One of the main ideas used in this work goes back to \cite{Galeati, FlaGalLuo},
see also \cite{FlaLuo, FlaGalLuo2}, but several other
aspects are new, first of all the way to overcome the difficulties due to the
boundary, but also the more quantitative presentation of the results, which
required new proofs.

Let us state the main result of this work. Let $J$ be a finite or countable
index set and $\left(  \mathbf{u}_{j}\left(  \mathbf{x}\right)  \right)
_{j\in J}$ be divergence free vector fields $\mathbf{u}_{j}:\overline
{D}\rightarrow\mathbb{R}^{d}$:%
\[
\mathbf{u}_{j}|_{\partial D}=0,\qquad\operatorname{div}\mathbf{u}_{j}=0
\]
with smoothness $\sum_{j\in J}\left\Vert \mathbf{u}_{j}\right\Vert
_{W^{1,2}\left(  D\right)  \cap C\left(  \overline{D}\right)  }^{2}<\infty$,
which in particular allows us to define the covariance matrix-valued function
$Q:\overline{D}\times\overline{D}\rightarrow\mathbb{R}^{d\times d}$%
\[
Q\left(  \mathbf{x},\mathbf{y}\right)  =\sum_{j\in J}\mathbf{u}_{j}\left(
\mathbf{x}\right)  \otimes\mathbf{u}_{j}\left(  \mathbf{y}\right),
\qquad\mathbf{x},\mathbf{y}\in\overline{D}.
\]
Associated to it define the bounded linear operator%
\[
\mathbb{Q}:L^{2}\big(  D;\mathbb{R}^{d}\big)  \rightarrow L^{2}\big(
D;\mathbb{R}^{d}\big)  ,\qquad\left(  \mathbb{Q}\mathbf{v}\right)  \left(
\mathbf{x}\right)  =\int_{D}Q\left(  \mathbf{x},\mathbf{y}\right)
\mathbf{v}\left(  \mathbf{y}\right)  d\mathbf{y}%
\]
and introduce two important quantities:%
\[
q\left(  \mathbf{x}\right)  :=\min_{\mathbf{\xi\neq0}}\frac{\mathbf{\xi}%
^{T}Q\left(  \mathbf{x},\mathbf{x}\right)  \mathbf{\xi}}{\mathbf{\xi}%
^{T}\mathbf{\xi}}, %
\]%
\[
\epsilon_{Q}:=\big\Vert \mathbb{Q}^{1/2} \big\Vert _{L^{2}\rightarrow L^{2}}^2 %
=\sup_{\mathbf{v\neq0}}\frac{\int_{D}\int_{D}\mathbf{v}\left(  \mathbf{x}%
\right)  ^{T}Q\left(  \mathbf{x},\mathbf{y}\right)  \mathbf{v}\left(
\mathbf{y}\right)  d\mathbf{x}d\mathbf{y}}{\int_{D}\mathbf{v}\left(
\mathbf{x}\right)  ^{T}\mathbf{v}\left(  \mathbf{x}\right)  d\mathbf{x}}.
\]
Denoted by $\big(  \Omega,\mathcal{F},(  \mathcal{F}_{t})_{t\geq0},\mathbb{P}\big)  $
a filtered probability space with expectation
$\mathbb{E}$, let $\big(  W_{t}^{j}\big)  _{j\in J}$ be a family of
independent Brownian motions; the generalized process
\[
\mathbf{u}_{Q}\left(  t,\mathbf{x}\right)  =\sum_{j\in J}\mathbf{u}_{j}\left(
\mathbf{x}\right)  \frac{dW_{t}^{j}}{dt}%
\]
is a white noise in time, divergence free, with space-covariance $Q\left(
\mathbf{x},\mathbf{y}\right)  $. We interpret the equation above as a
stochastic equation with Stratonovich noise (the precise interpretation is in
weak form with smooth test functions)%
\begin{equation} \label{stoch-heat-eq}
dT=\kappa\Delta Tdt+\sum_{j\in J}\mathbf{u}_{j}\cdot\mathbf{\nabla}T\circ
dW_{t}^{j}.
\end{equation}
Call $D\left(  A\right)  $ the space $W^{2,2}\left(  D\right)  \cap
W_{0}^{1,2}\left(  D\right)  $ where $W^{k,2}\left(  D\right)  $ are the
classical Sobolev spaces of square integrable $k$-times weakly differentiable
functions and $W_{0}^{1,2}\left(  D\right)  $ is the set of $W^{1,2}\left(
D\right)  $-functions equal to zero at the boundary. Define two linear
operators $A,A_{Q}:D\left(  A\right)  \rightarrow L^{2}\left(  D\right)  $ by
setting
\[
Af=\kappa\Delta f,\qquad A_{Q}f=\left(  \kappa\Delta+\mathcal{L}_{Q}\right)  f
\]
where%
\[
\left(  \mathcal{L}_{Q}f\right)  \left(  \mathbf{x}\right)  =\frac{1}{2}%
\sum_{\alpha,\beta=1}^{d}\partial_{\beta}\left(  Q_{\alpha\beta}\left(
\mathbf{x},\mathbf{x}\right)  \partial_{\alpha}f\left(  \mathbf{x}\right)
\right).
\]
Both operators $A,A_{Q}$ generate analytic semigroups \cite{Pazy} which we denote
by $e^{tA}$, $e^{tA_{Q}}$, $t\geq0$. The function $T_{Q}\left(  t,\mathbf{x}%
\right)  :=\big(  e^{tA_{Q}}T_{0}\big)  \left(  \mathbf{x}\right)  $ is the
solution of the modified heat equation
\[
\partial_{t}T_{Q}\left(  t,\mathbf{x}\right)  = \operatorname{div} \bigg[\Big(
\kappa I+ \frac12Q\left(  \mathbf{x},\mathbf{x}\right)  \Big)  \mathbf{\nabla}%
T_{Q}\left(  t,\mathbf{x}\right) \bigg]
\]
and thus, in view of the following result, we may call $Q\left(
\mathbf{x},\mathbf{x}\right)  $ the \textbf{eddy diffusion coefficient}.

We have denoted above by $\kappa\lambda_{D}$ the first eigenvalue of $-A$;
denote by $\lambda_{D,\kappa,Q}$ the first eigenvalue of $-A_{Q}$; a priori we
only know that $\lambda_{D,\kappa,Q}\geq\kappa\lambda_{D}$.

\begin{remark}
$\mathcal{L}_{Q}$ is a degenerate elliptic operator: since $\mathbf{u}%
_{j}|_{\partial D}=0$ we have also $Q|_{\partial D}=0$. Therefore it is not
clear a priori that $A_{Q}$ is more ``elliptic'' than $A$. However we shall
prove that $\lambda_{D,\kappa,Q}$ can be much larger than $\kappa\lambda_{D}$.
\end{remark}

Denote by $L_{\mathcal{F}_{0}}^{2}\big(  \Omega;L^{2}\left(  D\right)
\big)  $ the space of square integrable random variables with values in
$L^{2}\left(  D\right)  $, adapted to $\mathcal{F}_{0}$.

\begin{theorem}
\label{main thm}Assume $T_{0}\in L_{\mathcal{F}_{0}}^{2}\big(  \Omega
;L^{2}\left(  D\right)  \big)  $. Then, for every $\phi\in L^{\infty}\left(
D\right)  $,%
\[
\mathbb{E}\left[  \left(  \int_{D}\phi\left(  \mathbf{x}\right)  T\left(
t,\mathbf{x}\right)  d\mathbf{x}-\int_{D}\phi\left(  \mathbf{x}\right)  T_{Q}\left(
t,\mathbf{x}\right)  d\mathbf{x}\right)  ^{2}\right]  \leq\frac{\epsilon_{Q}}{2\kappa
}\mathbb{E}\left[  \left\Vert T_{0}\right\Vert _{L^{2}}^{2}\right]  \left\Vert
\phi\right\Vert _{\infty}^{2}.
\]
In particular, if $T_{0}\geq0$,
\[
\mathbb{E}\left[  \left(  \int_{D}\left\vert T\left(  t,\mathbf{x}\right)
\right\vert d\mathbf{x}\right)  ^{2}\right]  \leq\left(  \frac{\epsilon_{Q}}{\kappa
}+2\left\vert D\right\vert \exp\left(  -2\lambda_{D,\kappa,Q}t\right)
\right)  \mathbb{E}\left[  \left\Vert T_{0}\right\Vert _{L^{2}}^{2}\right]  .
\]

\end{theorem}

Here $|D|$ is the Lebesgue measure of $D$. This theorem shows (in a quantitative way) that decay is improved on
\textit{finite time intervals} $\left[  0,\tau\right]  $ if:

\begin{enumerate}
\item $\epsilon_{Q}$ is very small,

\item $\lambda_{D,\kappa,Q}\gg \kappa\lambda_{D}$.
\end{enumerate}

Denote by $D_{\delta}$ the set%
\[
D_{\delta}=\left\{  \mathbf{x}\in D:dist\left(  \mathbf{x},\partial D\right)
>\delta\right\}
\]
and assume%
\[
q\left(  \mathbf{x}\right)  \geq\sigma^{2}\quad\text{ in }D_{\delta}.
\]
For very general domains $D$, we have:

\begin{theorem}\label{theorem principal eigen asymptotic}
Let $D$ be an open, bounded, Lipschitz domain in $\mathbb{R}^{d}$.
Then, for any fixed $\kappa>0$, it holds
\[
\lim_{(\sigma,\delta)\rightarrow(+\infty,0)}\lambda_{D,\kappa,Q}=+\infty.
\]
\end{theorem}

Under more restrictive assumptions on the domain $D$ we may also provide the
following quantitative lower bound on $\lambda_{D,\kappa,Q}$:

\begin{theorem}\label{Thm eigen}
There exists a constant $C_{D,d}>0$ such that \
\[
\lambda_{D,\kappa,Q}\geq C_{D,d}\min\left(  \sigma^{2},\frac{\kappa}{\delta
}\right)
\]
for every $Q$ such that%
\[
q\left(  \mathbf{x}\right)  \geq\sigma^{2}\quad\text{ in }D_{\delta}.
\]
When $D$ is the unit ball, asymptotically as $\delta\rightarrow0$ one can take
$C_{D,d}=d/2$ and one also has $\lambda_{D,\kappa,Q}\geq\frac{\kappa d}%
{\kappa+\delta\sigma^{2}}\sigma^{2}$.
\end{theorem}

We prove all these claims in Section \ref{sect proofs}. The consequence of the
last two theorems is that $\lambda_{D,\kappa,Q}$ is large if the noise has a
large intensity function $q\left(  \mathbf{x}\right)  $, up to a small layer
around the boundary $\partial D$. Summarizing, the information given by
Theorems \ref{main thm}, \ref{theorem principal eigen asymptotic} and
\ref{Thm eigen} is that decay is improved on \textit{finite
time intervals} $\left[  0,\tau\right]  $ if:

\begin{enumerate}
\item $\epsilon_{Q}$ is very small,

\item $q\left(  \mathbf{x}\right)  $ is large, except for a small layer around
$\partial D$.
\end{enumerate}

The question then is: can we find a noise (namely a covariance function
$Q\left(  \mathbf{x},\mathbf{y}\right) )  $ with both properties, and possibly a
similarity with the statistics observed in turbulent fluids?

\begin{remark}
Notice that $\epsilon_{Q}$, by definition, is given by the operator norm
$\big\Vert \mathbb{Q}^{1/2}\big\Vert _{L^{2}\rightarrow L^{2}}^2$ and thus,
loosely speaking, it is related to the operator norm of $\mathbb{Q}$; and
$q\left(  \mathbf{x}\right)  $ is, loosely speaking, related to the trace of
the operator $\mathbb{Q}$:%
\[
Tr\left(  \mathbb{Q}\right)  =\int_{D}TrQ\left(  \mathbf{x},\mathbf{x}\right)
d\mathbf{x.}%
\]
The requirement that $\epsilon_{Q}$ is small and $q\left(  \mathbf{x}\right)
$ is large, heuristically translated into the requirements that the operator
norm of $\mathbb{Q}$ is small and the trace is large is not strange: many
operators have finite norm and infinite trace.
\end{remark}

First, we would like to explain an heuristic idea, which however we think of
relevance. We refer to a noise in full space; the translation in bounded
domain is a nontrivial issue under investigation. Consider the homogeneous
covariance ($Q\left(  \mathbf{x},\mathbf{y}\right)  =Q\left(  \mathbf{x}%
-\mathbf{y}\right)  $) of Kraichnan type%
\[
Q\left(  \mathbf{z}\right)  =\sigma^{2}k_{0}^{\zeta}\int_{k_{0}\leq\left\vert
\mathbf{k}\right\vert \leq k_{1}}\frac{1}{\left\vert \mathbf{k}\right\vert
^{d+\zeta}}e^{i\mathbf{k}\cdot\mathbf{z}} \bigg(I- \frac{\mathbf{k}\otimes\mathbf{k}%
}{\left\vert \mathbf{k}\right\vert ^{2}} \bigg) d\mathbf{k}. %
\]
There are two cases where conditions (i) and (ii) above are satisfied:

\begin{itemize}
\item if $\zeta>0$, $k_{1}=+\infty$, $\sigma^{2}$ large, and $k_{0}$ is so
large that $\sigma^{2}k_{0}^{-d}$ is small, then $q\left(  \mathbf{x}\right)
$ is large and $\epsilon_{Q}$ is small; recall \cite{Frisch} that K41 is $\zeta=\frac
{4}{3}$;

\item if $-d\leq\zeta\leq0$, $k_{0}=1$, $\sigma^{2}$ small, and $k_{1}$ is so
large that $\sigma^{2}\int_{1\leq k\leq k_{1}}\frac{1}{k^{\zeta+1}}dk$ is
large, then $q\left(  \mathbf{x}\right)  $ is large and $\epsilon_{Q}$ is
small; notice that $\zeta=-d$ is the case of white in space; and $\zeta=0$ is,
in dimension 2, the so called enstrophy measure.
\end{itemize}

In Section \ref{subsect Kraichnan}\ below we prove these claims. The previous
arguments require an excellent quantitative spectral knowledge which is not so
obvious in bounded domains; one could work with the eigenfunctions and
eigenvalues of Stokes operator, mimicking the previous claims, but it is
difficult to have explicit information to control the quantities. We have
preliminary results corresponding to the white noise case ($\zeta=-d$), not
reported here. Below, in Section \ref{subsect vortex noise}, we present a
different class of noise which, we believe, is new, suitable for bounded
domains and of interest in itself.

\section{Vortex patch noise}

The purpose of this section is the construction of a noise, in 2D, based on
the idea of vortex patches. The reader will recognize that a similar
construction can be done also in dimension 3 but the resulting objects look
artificial, since coherent vortex structures in 3D are closer to curves and
surfaces. But before, in order to identify a key step, we show why Kraichnan
noise works.

\subsection{Preliminaries on Kraichnan noise}\label{subsect Kraichnan}

Above we have claimed that Kraichnan noise produces large $q\left(
\mathbf{x}\right)  $ and small $\epsilon_{Q}$ under certain conditions. Let us
prove that claim because it requires a nontrivial argument in one step.
Missing that detail would spoil the understanding of the vortex patch noise
below. The control, for Kraichnan noise, on $q\left(  \mathbf{x}\right)  $ is
given by
\begin{align*}
\xi^{T}Q\left(  \mathbf{x},\mathbf{x}\right)  \xi &
=\xi^{T} Q\left(  0\right) \xi=\sigma^{2}k_{0}^{\zeta}%
\int_{k_{0}\leq\left\vert \mathbf{k}\right\vert \leq k_{1}}\frac{1}{\left\vert
\mathbf{k}\right\vert ^{d+\zeta}} \bigg(|\xi|^2- \frac{\left(  \mathbf{k}\cdot
\xi \right)  ^{2}}{\left\vert \mathbf{k}\right\vert ^{2}} \bigg) d\mathbf{k}\\
&  \geq\frac{3}{4} |\xi|^2 \sigma^{2}%
k_{0}^{\zeta}\int_{\substack{k_{0}\leq\left\vert \mathbf{k}\right\vert
\leq k_{1}\\ |\mathbf{k}\cdot\xi| \leq\left\vert \mathbf{k}\right\vert
\left\vert \xi \right\vert /2}} \frac{1}{\left\vert \mathbf{k}\right\vert
^{d+\zeta}}d\mathbf{k}\\
&  =\frac{3}{4} |\xi|^2 \sigma^{2}k_{0}^{\zeta
}C\int_{k_{0}}^{k_{1}}\frac{1}{r^{d+\zeta}}r^{d-1}dr=\frac{3}{4} |\xi|^2 \sigma^{2}C^{\prime}\left(  1-\left(
\frac{k_{0}}{k_{1}}\right)  ^{\zeta}\right)
\end{align*}
for suitable constants $C,C^{\prime}>0$. The control on $\epsilon_{Q}$, is
given by%
\begin{align*}
&  \int\int\mathbf{v}\left(  \mathbf{x}\right)  ^{T}Q\left(  \mathbf{x}%
,\mathbf{y}\right)  \mathbf{v}\left(  \mathbf{y}\right)  d\mathbf{x}%
d\mathbf{y}\\
&  =\sigma^{2}k_{0}^{\zeta}\int_{k_{0}\leq\left\vert \mathbf{k}\right\vert
\leq k_{1}}\frac{1}{\left\vert \mathbf{k}\right\vert ^{d+\zeta}}%
\bigg( | \widehat{\mathbf{v}}(\mathbf{k})|^2- \frac{\left\vert \mathbf{k}\cdot \widehat{\mathbf{v}}\left(  \mathbf{k}\right)  \right\vert
^{2}}{\left\vert \mathbf{k}\right\vert ^{2}%
} \bigg) d\mathbf{k}\\
&  \leq\sigma^{2}k_{0}^{-d}\int_{k_{0}\leq\left\vert \mathbf{k}\right\vert
\leq k_{1}}\left\vert \widehat{\mathbf{v}}\left(  \mathbf{k}\right)
\right\vert ^{2}d\mathbf{k}\leq\sigma^{2}k_{0}^{-d}\left\Vert \mathbf{v}%
\right\Vert _{L^{2}}^{2}.
\end{align*}
It is here that one step must be performed in the right way. If we just estimate from above as%
\[
\int\int\mathbf{v}\left(  \mathbf{x}\right)  ^{T}Q\left(  \mathbf{x}%
,\mathbf{y}\right)  \mathbf{v}\left(  \mathbf{y}\right)  d\mathbf{x}%
d\mathbf{y}\leq\int\int\sigma^{2}k_{0}^{\zeta}\int_{k_{0}\leq\left\vert
\mathbf{k}\right\vert \leq k_{1}}\frac{1}{\left\vert \mathbf{k}\right\vert
^{d+\zeta}}\left\vert \mathbf{v}\left(  \mathbf{x}\right)  \right\vert
\left\vert \mathbf{v}\left(  \mathbf{y}\right)  \right\vert d\mathbf{k}%
d\mathbf{x}d\mathbf{y}%
\]
then, first, we are in trouble since the $L^{1}$ norm of $\mathbf{v}$\ is
difficult to estimate. Second, even if the space domain is a Torus (in this
case the integral over wave numbers is a series) we would end-up with an
estimate of the form%
\[
\leq\sigma^{2}k_{0}^{\zeta}\left\Vert \mathbf{v}\right\Vert _{L^{2}}^{2}%
\sum_{k_{0} \leq \left\vert \mathbf{k}\right\vert \leq k_{1}}\frac{1}{\left\vert
\mathbf{k}\right\vert ^{d+\zeta}} \leq C\sigma^{2}\left\Vert
\mathbf{v}\right\Vert _{L^{2}}^{2}\left(  1-\left(  \frac{k_{0}}{k_{1}%
}\right)  ^{\zeta}\right)
\]
which is not sufficient. The result would be that there is no difference in
estimating the norm or the trace. The key is using the presence of an
orthonormal family of functions (here $e^{i\mathbf{k}\cdot\mathbf{z}}$).

\subsection{The vortex noise in 2D}\label{subsect vortex noise}

Thus consider $d=2$ and assume that $D$ is a smooth bounded connected open
domain. We are going to describe a noise of the form $\sum_{j\in J}%
\mathbf{u}_{j}\left(  \mathbf{x}\right)  dW_{t}^{j}$ with
\[
\mathbf{u}_{j}\left(  \mathbf{x}\right)  =\mathbf{w}_{r}\left(  \mathbf{x}%
-\mathbf{x}_{j}\right)  ,\qquad\mathbf{w}_{r}\left(  \mathbf{x}\right)
=r^{-1}\mathbf{w}\left(  \frac{\mathbf{x}}{r}\right)
\]
for suitable $r$ and $\mathbf{w}$. The ingredients are therefore the points
$\mathbf{x}_{j}$, called the \textquotedblleft centers\textquotedblright\ of
the vortex blobs below, and a vector field $\mathbf{w}$.

\subsubsection{The centers of the vortex blobs}

Given a positive integer $N$ such that $\frac{1}{N}\leq\delta$, consider the
set $\Lambda_{N}$ of all points of $D_{\delta}$ having coordinates of the form
$\big(\frac{k}{N},\frac{h}{N}\big)$ with $k,h\in\mathbb{Z}$. For the purpose
of the example developed here, the centers $\mathbf{x}_{j}$ of the blobs will
be taken equal to the points of $\Lambda_{N}$; with some effort one can
generalize to more flexible distributions of points, also random.

The index set $J$ will be $\Lambda_{N}$ itself and points of $\Lambda_{N}$
will be denoted by $\mathbf{z}$. Notations below in this section will adapt to
this choice; for instance we write the noise in the form%
\[
\Gamma \sum_{\mathbf{z}\in\Lambda_{N}}\mathbf{w}_{r}\left(  \mathbf{x}-\mathbf{z}%
\right)  dW_{t}^{\mathbf{z}}.
\]
We have%
\[
\min_{\mathbf{z}_{1}\neq\mathbf{z}_{2}\in\Lambda_{N}}\left\vert \mathbf{z}%
_{1}-\mathbf{z}_{2}\right\vert =\frac{1}{N},\qquad\min_{\mathbf{z}\in
\Lambda_{N}}d\left(  \mathbf{z},\partial D\right)  \geq\delta.
\]

Given a positive integer $M$ (in the sequel $M$ will be finite, while
$N\rightarrow\infty$), the set $\Lambda_{N}$ is decomposed as the disjoint
union of the sets%
\[
\Lambda_{N}=\bigcup_{(k_{0},h_{0})\in\{0,1,...,M-1\}^{2}}\Lambda_{N}^{\left(
M,k_{0},h_{0}\right)  }%
\]
defined as follows: the points $\big(\frac{k}{N},\frac{h}{N}\big)$ of
$\Lambda_{N}^{\left(  M,k_{0},h_{0}\right)  }$ have the property that
$k=Mn+k_{0}$, $h=Mm+h_{0}$, with $n,m\in\mathbb{Z}$. Therefore%
\[
\min_{\mathbf{z}_{1}\neq\mathbf{z}_{2}\in\Lambda_{N}^{\left(  M,k_{0}%
,h_{0}\right)  }}\left\vert \mathbf{z}_{1}-\mathbf{z}_{2}\right\vert =\frac
{M}{N}%
\]
for each $\left(  k_{0},h_{0}\right)  \in\left\{  0,1,...,M-1\right\}  ^{2}$.

\subsubsection{The vector field $\mathbf{w}$}

The construction of vector field $\mathbf{w}$ requires some care. First, in
order to have that $\sum_{\mathbf{z}\in\Lambda_{N}}\mathbf{w}_{r}\left(
\mathbf{x}-\mathbf{z}\right)  dW_{t}^{\mathbf{z}}$ is an admissible noise for
our investigation, we need that each $\mathbf{u}_{\mathbf{z}}\left(
\mathbf{x}\right)  :=\mathbf{w}_{r}\left(  \mathbf{x}-\mathbf{z}\right)  $ is
divergence free, smooth enough and zero at $\partial D$. Therefore we need
$\operatorname{div}\mathbf{w}=0$, $\mathbf{w}$\ smooth enough; and we look for
a vector field with compact support, say in the closed ball $\overline
{B\left(  0,1\right)  }$, so that for $r\in\left(  0,\delta\right)  $ and
$\mathbf{z\in}\Lambda_{N}\subset D_{\delta}$ the rescaled and shifted vector
field $\mathbf{w}_{r}\left(  \mathbf{x}-\mathbf{z}\right)  $ is zero on
$\partial D$. Moreover, we need other two properties.

One is that $\mathbf{w}\left(  \mathbf{x}\right)  $ is close to $\frac{1}%
{2\pi}\frac{\mathbf{x}^{\perp}}{\left\vert \mathbf{x}\right\vert ^{2}}$ near
$\mathbf{x}=0$; this is central to the proof that the function $q\left(
\mathbf{x}\right)  $ is large. The other is that the vector fields
$\mathbf{w}_{r}\left(  \mathbf{x}-\mathbf{z}\right)  $ are (up to the constant
$\int\left\vert \mathbf{w}\left(  \mathbf{x}\right)  \right\vert
^{2}d\mathbf{x}$, which is not zero since $\mathbf{w}$ is close to $\frac
{1}{2\pi}\frac{\mathbf{x}^{\perp}}{\left\vert \mathbf{x}\right\vert ^{2}}$
near $\mathbf{x}=0$) ``almost'' orthonormal in $L^{2}$, which is guaranteed by
the fact that the supports are ``almost'' disjoint. To be precise, if we take
truly disjoint supports, then the action of $\mathbf{w}_{r}\left(
\mathbf{x}-\mathbf{z}\right)  $ does not cover the full set $D_{2\delta}$:
there are intermediate zones between the supports, where $\mathbf{w}%
_{r}\left(  \mathbf{x}-\mathbf{z}\right)  $ does not move space points and
this is in contrast with the requirement that $q\left(  \mathbf{x}\right)  $
should be large everywhere in $D_{2\delta}$. This is why we have introduced
$M$ and the sets $\Lambda_{N}^{\left(  M,k_{0},h_{0}\right)  }$ above: inside
each one of these classes the supports will be disjoint and this is sufficient
for our estimates; in order to have the supports disjoint for elements of
$\Lambda_{N}^{\left(  M,k_{0},h_{0}\right)  }$ we ask $r\leq\frac{M}{2N}$.

Therefore, summarizing, we look for a vector field $\mathbf{w}$, defined on
$\mathbb{R}^{2}$, smooth, with compact support in $\overline{B\left(
0,1\right)  }$, $\operatorname{div}\mathbf{w}=0$, close to $\frac{1}{2\pi
}\frac{\mathbf{x}^{\perp}}{\left\vert \mathbf{x}\right\vert ^{2}}$ near
$\mathbf{x}=0$. We construct it as%
\[
\mathbf{w}=\mathbf{\nabla}^{\perp}\psi
\]
so that it is divergence free. Thus we look for a smooth function $\psi$ on
$\mathbb{R}^{2}$, compactly supported in $\overline{B\left(  0,1\right)  }$,
close to $\frac{1}{2\pi}\log\left\vert \mathbf{x}\right\vert $ near
$\mathbf{x}=0$. Such function exists and can be constructed in several ways.

Let $\psi_{0}\in C^{\infty}\big(\mathbb{R}^{2}\backslash\left\{  0\right\}
\big)$ be a radial function such that
\[
\psi_{0}\left(  \mathbf{x}\right)  =\frac{1}{2\pi}\log\left\vert
\mathbf{x}\right\vert \text{ for }\left\vert \mathbf{x}\right\vert \leq
\frac{1}{3}\text{ and }\psi_{0}\left(  \mathbf{x}\right)  =0\text{ for
}\left\vert \mathbf{x}\right\vert >\frac{2}{3}.
\]
Let $f\in C^{\infty}\big(\mathbb{R}^{2}\big)$ be a probability density
function with support in $B\left(  0,1\right)  $. Given $\epsilon>0$ small (at
least $\epsilon<\frac{1}{6}$), define%
\[
f_{\epsilon}\left(  \mathbf{x}\right)  =\epsilon^{-2}f\left(  \frac
{\mathbf{x}}{\epsilon}\right)  ,\qquad\psi\left(  \mathbf{x}\right)
=\int_{\mathbb{R}^{2}}\psi_{0}\left(  \mathbf{x}-\mathbf{y}\right)
f_{\epsilon}\left(  \mathbf{y}\right)  d\mathbf{y}.
\]
This function satisfies our requirements: its support is in $\overline
{B\left(  0,1\right)  }$, it is smooth everywhere and, if we take $\epsilon$
small, it is close to $\psi_{0}$ which is equal to $\frac{1}{2\pi}%
\log\left\vert \mathbf{x}\right\vert $ near $\mathbf{x}=0$. The corresponding
vector field $\mathbf{w}=\mathbf{\nabla}^{\perp}\psi$ has the required properties.

Therefore, if $\left\vert \mathbf{x}\right\vert \leq\frac{1}{6}$ and
$\epsilon<\frac{1}{6}$ (so that the support of $f_{\epsilon}$ is in $B\left(
0,\frac{1}{6}\right)  $) we have\textbf{ }%
\[
\mathbf{w}\left(  \mathbf{x}\right)  =\int_{\mathbb{R}^{2}}\mathbf{\nabla
}^{\perp}\psi_{0}\left(  \mathbf{x}-\mathbf{y}\right)  f_{\epsilon}\left(
\mathbf{y}\right)  d\mathbf{y}=\int_{\mathbb{R}^{2}}\frac{1}{2\pi}%
\frac{\left(  \mathbf{x}-\mathbf{y}\right)  ^{\perp}}{\left\vert
\mathbf{x}-\mathbf{y}\right\vert ^{2}}f_{\epsilon}\left(  \mathbf{y}\right)
d\mathbf{y}, %
\]%
\[
\mathbf{w}_{r}\left(  \mathbf{x}\right)  =\frac{1}{2\pi r}\int_{\mathbb{R}%
^{2}}\frac{\left(  \mathbf{x}/r-\mathbf{y}\right)  ^{\perp}}{\left\vert
\mathbf{x}/r-\mathbf{y}\right\vert ^{2}}f_{\epsilon}\left(  \mathbf{y}\right)
d\mathbf{y}=\frac{1}{2\pi}\int_{\mathbb{R}^{2}}\frac{\left(  \mathbf{x}%
-\mathbf{y}\right)  ^{\perp}}{\left\vert \mathbf{x}-\mathbf{y}\right\vert
^{2}}\left(  \epsilon r\right)  ^{-2}f\left(  \mathbf{y}/\left(  \epsilon
r\right)  \right)  d\mathbf{y.}%
\]

\subsubsection{Estimates on $q\left(  \mathbf{x}\right)  $ and $\epsilon_{Q}%
$.}

We now check that, with proper choices of the parameters, the noise
$\Gamma \sum_{\mathbf{z}\in\Lambda_{N}}\mathbf{w}_{r}\left(  \mathbf{x}%
-\mathbf{z}\right)  dW_{t}^{\mathbf{z}}$ with $\mathbf{w}_{r}\left(
\mathbf{x}\right)  =r^{-1}\mathbf{w}\left(  \frac{\mathbf{x}}{r}\right)  $ has
large $q\left(  \mathbf{x}\right)  $ and small $\epsilon_{Q}$.

We choose $r$ with more than one constraint. We have already assumed above
\[
r\leq\frac{M}{2N},\qquad r\leq\delta.
\]
The first inequality implies that the supports of $\mathbf{w}_{r}\left(
\mathbf{x}-\mathbf{z}\right)  $ are disjoint for $\mathbf{z}$ in the same
subset $\Lambda_{N}^{\left(  M,k_{0},h_{0}\right)  }$. The second inequality
implies that they are zero at the boundary of $D$.

The covariance of this noise is%
\[
Q\left(  \mathbf{x},\mathbf{y}\right)  =\Gamma^{2}\sum_{\mathbf{z}\in
\Lambda_{N}}\mathbf{w}_{r}\left(  \mathbf{x}-\mathbf{z}\right)  \otimes
\mathbf{w}_{r}\left(  \mathbf{y}-\mathbf{z}\right)  .
\]
We therefore have, for the estimate of $\epsilon_{Q}$,
\begin{align*}
&  \int\int\mathbf{v}\left(  \mathbf{x}\right)  ^{T}Q\left(  \mathbf{x}%
,\mathbf{y}\right)  \mathbf{v}\left(  \mathbf{y}\right)  d\mathbf{x}%
d\mathbf{y}=\Gamma^{2}\sum_{\mathbf{z}\in\Lambda_{N}}\left(  \int%
\mathbf{w}_{r}\left(  \mathbf{x}-\mathbf{z}\right)  \cdot\mathbf{v}\left(
\mathbf{x}\right)  d\mathbf{x}\right)  ^{2}\\
&  =\left\Vert \mathbf{w}\right\Vert _{L^{2}}^{2}\Gamma^{2}\sum_{\left(
k_{0},h_{0}\right)  \in\left\{  0,1,...,M-1\right\}  ^{2}}\sum_{\mathbf{z}%
\in\Lambda_{N}^{\left(  M,k_{0},h_{0}\right)  }}\left(  \int\frac
{\mathbf{w}_{r}\left(  \mathbf{x}-\mathbf{z}\right)  }{\left\Vert
\mathbf{w}\right\Vert _{L^{2}}}\cdot\mathbf{v}\left(  \mathbf{x}\right)
d\mathbf{x}\right)  ^{2}\\
&  \leq M^{2}\left\Vert \mathbf{w}\right\Vert _{L^{2}}^{2}\Gamma^{2}\left\Vert
\mathbf{v}\right\Vert _{L^{2}}^{2}.
\end{align*}
We have used a basic property, similarly to the most important step in the
verification done above for Kraichnan noise: the family $\Big\{\frac
{\mathbf{w}_{r}\left(  \mathbf{x}-\mathbf{z}\right)  }{\left\Vert
\mathbf{w}\right\Vert _{L^{2}}}\Big\}_{\mathbf{z}\in\Lambda_{N}^{\left(
M,k_{0},h_{0}\right)  }}$ is orthonormal (not complete), because of the
disjoint supports and the property $\int\left\vert \mathbf{w}_{r}\left(
\mathbf{x}\right)  \right\vert ^{2}d\mathbf{x}=\left\Vert \mathbf{w}%
\right\Vert _{L^{2}}^{2}$. One can easily check that
\[
\left\Vert \mathbf{w}\right\Vert _{L^{2}}^{2}\leq C\log\frac{1}{\epsilon}
\]
and therefore, taking $\epsilon=\frac1N$ leads to
\[
\epsilon_{Q}\leq M^{2}\Gamma^{2}C\log N
\]
which is small if, given $N$, $\Gamma$ is small enough.

Concerning $q\left(  \mathbf{x}\right)  $, we have, for every $\mathbf{x}\in
D$ and every unitary vector $\mathbf{v}\in\mathbb{R}^{2}$,%
\[
\mathbf{v}^{T}Q\left(  \mathbf{x},\mathbf{x}\right)  \mathbf{v}=\Gamma^{2}%
\sum_{\mathbf{z}\in\Lambda_{N}}\left(  \mathbf{w}_{r}\left(  \mathbf{x}%
-\mathbf{z}\right)  \cdot\mathbf{v}\right)  ^{2}.
\]
Now, consider a point $\mathbf{x}\in D_{2\delta}$. If $N$ is large enough with
respect to the curvature of $\partial D$ near $\mathbf{x}$, we may find
$\mathbf{z}\in\Lambda_{N}$ close to $\mathbf{x}$, precisely with $\frac1{2N} \leq \left\vert
\mathbf{x}-\mathbf{z}\right\vert <\frac{2}{N}$, such that
\[
\left\vert \mathbf{v}\cdot\frac{\left(  \mathbf{x}-\mathbf{z}\right)  ^{\perp
}}{\left\vert \mathbf{x}-\mathbf{z}\right\vert }\right\vert \geq\frac{1}{4}.
\]
Then, if $\left\vert \frac{\mathbf{x}-\mathbf{z}}{r}\right\vert \leq\frac
{1}{6}$, which is true if $\frac{2}{rN}\leq\frac{1}{6}$, namely $r\geq
\frac{12}{N}$,
\[
\left\vert \mathbf{w}_{r}\left(  \mathbf{x}-\mathbf{z}\right)  \cdot
\mathbf{v}\right\vert =\left\vert \frac{1}{2\pi}\int_{\mathbb{R}^{2}%
}\mathbf{v}\cdot\frac{\left(  \mathbf{x}-\mathbf{z}-\mathbf{y}\right)
^{\perp}}{\left\vert \mathbf{x}-\mathbf{z}-\mathbf{y}\right\vert ^{2}}\left(
\epsilon r\right)  ^{-2}f\left(  \mathbf{y}/\left(  \epsilon r\right)
\right)  d\mathbf{y}\right\vert .
\]
The constraints $r\leq\frac{M}{2N}$, $r\leq\delta$, $r\geq\frac{12}{N}$
are all satisfiable if we take $M>24$ and $N$ large enough; of course we may
reduce quantitatively the constraint $M>24$ by different choices of some
parameters above. Recalling that $\epsilon=\frac1N$; in the above integral,
we have $|\mathbf{y}|\leq \epsilon r \sim \frac1{N^2}$ which means that $\mathbf{y}$ is
an infinitesimal perturbation of $\mathbf{x}-\mathbf{z}$ due to $|\mathbf{x}-\mathbf{z}|\sim \frac1N$.
Thus, for $N$ big enough the last integral is bounded below by%
\[
\geq\frac{1}{2\pi}\frac{1}{8}N=\frac{N}{16\pi}.
\]
It follows%
\[
q\left(  \mathbf{x}\right)  \geq\frac{\Gamma^{2}N}{16\pi}.
\]
Therefore we may choose $N$ and $\Gamma$ so that $\epsilon_{Q}$ is small as we
want and $q\left(  \mathbf{x}\right)  $, on $D_{2\delta}$, is large as we
want.

\section{Proofs}\label{sect proofs}

For reasons of space, we omit some secondary details in the following proofs;
for instance we do not write explicitly the definition of solution, the proof
that energy and maximum principle estimates are satisfied, the proof that we
may pass from the weak to the mild formulation.

\subsection{Proof of Theorem \ref{main thm}}

The first key ingredient is the reformulation of the Stratonovich equation in
It\^{o} form%
\[
d_{t}T=\big(  \kappa\Delta T+\widetilde{\mathcal{L}}_{Q}T\big)
dt+\sum_{j\in J}\mathbf{u}_{j}\cdot\mathbf{\nabla}TdW_{t}^{j}, %
\]
where
\[
\big(\widetilde{\mathcal{L}}_{Q}T\big)\left(  \mathbf{x}\right)  :=\frac{1}%
{2}\sum_{j\in J}\mathbf{u}_{j}\left(  \mathbf{x}\right)  \cdot\mathbf{\nabla
}\left(  \mathbf{u}_{j}\left(  \mathbf{x}\right)  \cdot\mathbf{\nabla}T\left(
\mathbf{x}\right)  \right)  .
\]
One has%
\[
\widetilde{\mathcal{L}}_{Q}=\mathcal{L}_{Q}.
\]
This is a well known fact, see for instance \cite{CoghiFla}; indeed%

\[
\widetilde{\mathcal{L}}_{Q}T=\frac{1}{2}\sum_{j\in J}\sum_{\alpha,\beta=1}%
^{d}u_{j}^{\alpha}\partial_{\alpha}u_{j}^{\beta}\partial_{\beta}T+\frac{1}%
{2}\sum_{j\in J}\sum_{\alpha,\beta=1}^{d}u_{j}^{\alpha}u_{j}^{\beta}%
\partial_{\alpha}\partial_{\beta}T.
\]
The second sum is equal to $\frac{1}{2}\sum_{\alpha,\beta=1}^{d}Q_{\alpha
\beta}\left(  \mathbf{x},\mathbf{x}\right)  \partial_{\alpha}\partial_{\beta
}T$. The first one, due to the property $\operatorname{div}\mathbf{u}_{j}=0$,
is equal to
\[
\frac{1}{2}\sum_{j\in J}\sum_{\alpha,\beta=1}^{d}\partial_{\alpha}\left(
u_{j}^{\alpha}u_{j}^{\beta}\right)  \partial_{\beta}T=\frac{1}{2}\sum
_{\alpha,\beta=1}^{d}\partial_{\alpha}Q_{\alpha\beta}\left(  \mathbf{x}%
,\mathbf{x}\right)  \partial_{\beta}T
\]
where we have also used the assumptions of uniform convergence of the series
of the derivatives.

From the previous facts we have%
\[
d_{t}\left(  T-T_{Q}\right)  =\left(  \kappa\Delta+\mathcal{L}_{Q}\right)
\left(  T-T_{Q}\right)  dt+\sum_{j\in J}\mathbf{u}_{j}\cdot\mathbf{\nabla
}TdW_{t}^{j}.
\]
The mild formulation of this identity, furthermore applied in a weak sense to
a smooth test function $\phi$ with compact support in $D$, is:%
\[
\left\langle \phi,T\left(  t\right)  -T_{Q}\left(  t\right)  \right\rangle
=\sum_{j\in J}\int_{0}^{t}\left\langle e^{\left(  t-s\right)  A_{Q}}%
\phi,\mathbf{u}_{j}\cdot\mathbf{\nabla}T\left(  s\right)  \right\rangle
dW_{s}^{j}%
\]
where $\langle\cdot, \cdot\rangle$ is the inner product in $L^2(D)$ and
we have also used the fact that the semigroup $e^{\left(  t-s\right)
A_{Q}}$ is self adjoint. By the isometry formula for It\^{o} integrals,%
\[
\mathbb{E}\left[  \left\langle \phi,T\left(  t\right)  -T_{Q}\left(  t\right)
\right\rangle ^{2}\right]  =\sum_{j\in J}\int_{0}^{t}\mathbb{E}\left[
\left\langle e^{\left(  t-s\right)  A_{Q}}\phi,\mathbf{u}_{j}\cdot
\mathbf{\nabla}T\left(  s\right)  \right\rangle ^{2}\right]  ds.
\]
We have (we write $T_{s}\left(  \mathbf{x}\right)  $ for $T\left(
s,\mathbf{x}\right)  $ and $\phi_{t,s}\left(  \mathbf{x}\right)  $ for
$\big(  e^{\left(  t-s\right)  A_{Q}}\phi \big)  \left(  \mathbf{x}\right)
$ to shorten notations)
\begin{align*}
&  \sum_{j\in J}\left\langle e^{\left(  t-s\right)  A_{Q}}\phi,\mathbf{u}%
_{j}\cdot\mathbf{\nabla}T_{s}\right\rangle ^{2}\\
&  =\sum_{\alpha,\beta=1}^{d}\int_{D}\int_{D}\phi_{t,s}\left(  \mathbf{x}%
\right)  \phi_{t,s}\left(  \mathbf{y}\right)  Q_{\alpha\beta}\left(
\mathbf{x,y}\right)  \partial_{\alpha}T_{s}\left(  \mathbf{x}\right)
\partial_{\beta}T_{s}\left(  \mathbf{y}\right)  d\mathbf{x}d\mathbf{y.}%
\end{align*}
The semigroup $e^{tA_{Q}}$ satisfies the Maximum Principle, namely $\big\Vert
e^{tA_{Q}}\phi \big\Vert _{\infty}\leq\left\Vert \phi\right\Vert _{\infty}$.
Hence, recalling the definition of $\mathbb{Q}$ and $\epsilon_{Q}$, %
\begin{align*}
\sum_{j\in J}\left\langle e^{\left(  t-s\right)  A_{Q}}\phi,\mathbf{u}%
_{j}\cdot\mathbf{\nabla}T_{s}\right\rangle ^{2} & \leq \epsilon_{Q}
\int_{D} \big|\phi_{t,s}(\mathbf{x})\, \nabla T_s(\mathbf{x}) \big|^2 d\mathbf{x} \\
&  \leq\left\Vert \phi\right\Vert _{\infty}^{2}\epsilon_{Q}\int_{D}\left\vert
\mathbf{\nabla}T\left(  s,\mathbf{x}\right)  \right\vert ^{2}d\mathbf{x}. %
\end{align*}
Moreover, for the
original stochastic equation \eqref{stoch-heat-eq} we have the inequality%
\[
\int_{0}^{\infty}\int_{D}\left\vert \mathbf{\nabla}T\left(  t,\mathbf{x}%
\right)  \right\vert ^{2}d\mathbf{x}dt\leq\frac{1}{2\kappa}\int_{D}T_{0}%
^{2}\left(  \mathbf{x}\right)  d\mathbf{x.}%
\]
Together they imply
\[
\mathbb{E}\left[  \left\langle \phi,T\left(  t\right)  -T_{Q}\left(  t\right)
\right\rangle ^{2}\right]  \leq\frac{\epsilon_{Q}}{2\kappa}\mathbb{E}\left[
\left\Vert T_{0}\right\Vert _{L^{2}}^{2}\right]  \left\Vert \phi\right\Vert
_{\infty}^{2}.
\]

If $T_{0}\geq0$, then both $T\left(  t\right)  $ and $T_{Q}\left(  t\right)  $
are nonnegetive. Choose a sequence $\phi_{n}$ converging to 1 in $D$. We deduce
\[
\mathbb{E}\left[  \left(  \int_{D}T\left(  t,\mathbf{x}\right)  d\mathbf{x}%
-\left\langle 1,e^{tA_{Q}}T_{0}\right\rangle \right)  ^{2}\right]
\leq\frac{\epsilon_{Q}}{2\kappa}\mathbb{E}\left[  \left\Vert T_{0}\right\Vert
_{L^{2}}^{2}\right]  .
\]
It implies%
\begin{align*}
\mathbb{E}\left[  \left(  \int_{D}\left\vert T\left(  t,\mathbf{x}\right)
\right\vert d\mathbf{x}\right)  ^{2}\right]   &  \leq2\frac{\epsilon_{Q}%
}{2\kappa}\mathbb{E}\left[  \left\Vert T_{0}\right\Vert_{L^2} ^{2}\right]
+2\mathbb{E}\left[  \left\langle 1,e^{tA_{Q}}T_{0}\right\rangle _{L^{2}}%
^{2}\right]  \\
&  \leq\frac{\epsilon_{Q}}{\kappa}\mathbb{E}\left[  \left\Vert T_{0}%
\right\Vert _{L^{2}}^{2}\right]  +2\left\vert D\right\vert \mathbb{E}\left[
\big\Vert e^{tA_{Q}}T_{0}\big\Vert _{L^{2}}^{2}\right]  \\
&  \leq\left(  \frac{\epsilon_{Q}}{\kappa}+2\left\vert D\right\vert
\exp\left(  -2\lambda_{D,\kappa,Q}t\right)  \right)  \mathbb{E}\left[
\left\Vert T_{0}\right\Vert _{L^{2}}^{2}\right]  .
\end{align*}

\subsection{Proof of Theorems \ref{theorem principal eigen asymptotic} and \ref{Thm eigen}}

\subsubsection{Proof of Theorem \ref{theorem principal eigen asymptotic}}

We use the variational characterization of $\lambda_{D,\kappa,Q}$ given by
\[
\lambda_{D,\kappa,Q}=\inf_{T\in W_{0}^{1,2}\left(  D\right)  :\int_{D}%
T^{2}d\mathbf{x} =1}\int_{D}\sum_{\alpha,\beta=1}^{d}\left(  \kappa\delta_{\alpha\beta
}+Q_{\alpha\beta}\left(  \mathbf{x,x}\right)  \right)  \partial_{\alpha
}T\left(  \mathbf{x}\right)  \partial_{\beta}T\left(  \mathbf{x}\right)
d\mathbf{x}.
\]
We have $\lambda_{D,\kappa,Q}\geq\lambda_{\kappa,\sigma,\delta}$ where%
\[
\lambda_{\kappa,\sigma,\delta}:=\inf_{T\in W_{0}^{1,2}\left(  D\right)
:\int_{D}T^{2}d\mathbf{x}=1}\int_{D}\left(  \kappa+\sigma^{2}\cdot1_{D_{\delta}}\left(
\mathbf{x}\right)  \right)  |\mathbf{\nabla}T\left(  \mathbf{x}\right)
|^{2}d\mathbf{x}.
\]
We want to prove that
\[
\lim_{(\sigma,\delta)\rightarrow(+\infty,0)}\lambda_{\kappa,\sigma,\delta
}=+\infty.
\]
Suppose this is not true, then we can find $C>0$ and a sequence $(\sigma
_{n},\delta_{n})\rightarrow(+\infty,0)$ such that $\lambda_{n}:=\lambda
_{\kappa,\sigma_{n},\delta_{n}}\leq C$; this implies that we can find a
sequence $T_{n}\in W_{0}^{1,2}(D)$ such that $\Vert T_{n}\Vert_{L^{2}}=1$ and
\[
\int_{D}(\kappa+\sigma_{n}^{2}\cdot1_{D_{\delta_{n}}}(\mathbf{x}%
))|\mathbf{\nabla}T_{n}(\mathbf{x})|^{2}d\mathbf{x}=\lambda_{n}\leq
C\quad\forall\,n\in\mathbb{N}.
\]
We deduce as a consequence that $\int_{D}|\mathbf{\nabla}T_{n}|^{2}%
d\mathbf{x}\leq\kappa^{-1}C$ and the sequence $\{T_{n}\}_{n}$ is bounded in
$W_{0}^{1,2}(D)$; by Rellich-Kondrakhov compactness theorem for $W_{0}%
^{1,2}(D)$, we can extract a (not relabelled) subsequence such that
$T_{n}\rightarrow T$ strongly in $L^{2}(D)$ and $\mathbf{\nabla}%
T_{n}\rightarrow\mathbf{\nabla}T$ weakly in $L^{2}(D)$ for a suitable $T\in
W_{0}^{1,2}(D)$. On the other hand,
\[
\int_{D_{\delta_{n}}}|\mathbf{\nabla}T_{n}(\mathbf{x})|^{2}d\mathbf{x}%
\leq\frac{C}{\sigma_{n}^{2}}\rightarrow0
\]
which together with $D_{\varepsilon}\subset D_{\delta_{n}}$ for $n$ large
enough implies that $\mathbf{\nabla}T(\mathbf{x})=0$ for a.e. $x\in
D_{\varepsilon}$ and for any $\varepsilon>0$. Overall this implies that $\Vert
T\Vert_{L^{2}}=1$ and $\mathbf{\nabla}T=0$, thus $T$ is a constant function
which is $0$ at the boundary $\partial D$, giving a contradiction.

\subsubsection{Preparation to the proof of Theorem \ref{Thm eigen}}

We give the proof of Theorem \textit{\ref{Thm eigen}} only in the case of the
ball $D=B\left(  0,1\right)  $. The case of a star-shaped domain with smooth
boundary can be reduced to the ball by relatively easy arguments. We think
that the result is true for much more general domains but the details are
outside the scope of this work.

Therefore now we have (with the notations of the previous section)%
\[
\lambda_{\kappa,\sigma,\delta}=\inf_{T\in W_{0}^{1,2}\left(  B\left(
0,1\right)  \right)  :\int_{B\left(  0,1\right)  }T^{2} d\mathbf{x}=1}\int_{B\left(
0,1\right)  }\left(  \kappa+\sigma^{2}\cdot1_{B\left(  0,1-\delta\right)
}\left(  \mathbf{x}\right)  \right)  |\mathbf{\nabla}T\left(  \mathbf{x}%
\right)  |^{2}d\mathbf{x}.
\]
Classical facts guarantee that there is a unique minimizer for the variational
problem which defines $\lambda_{\kappa,\sigma,\delta}$, and it is
non-negative. Denote it by $T_{\kappa,\sigma,\delta}$. Since the functional is
invariant by rotation, uniqueness implies that also the minimizer is invariant
by rotation. Then%
\[
T_{\kappa,\sigma,\delta}\left(  \mathbf{x}\right)  =f_{\kappa,\sigma,\delta
}\left(  \left\vert \mathbf{x}\right\vert \right)
\]
for some function $f_{\kappa,\sigma,\delta }\in W^{1,2}\left(  0,1\right)  $. Called $\omega_{d}$ the
surface of the unit sphere in $\mathbb{R}^{d}$, we have $\lambda
_{\kappa,\sigma,\delta}=\omega_{d}\inf J\left(  f\right)  $,%
\begin{equation}\label{J-functional}
J\left(  f\right)  =\kappa\int_{0}^{1}f^{\prime}\left(  r\right)  ^{2}%
r^{d-1}dr+\sigma^{2}\int_{0}^{1-\delta}f^{\prime}\left(  r\right)  ^{2}%
r^{d-1}dr
\end{equation}
the infimum being taken over all $f\in W^{1,2}\left(  0,1\right)  $ such that
$f\left(  1\right)  =0$ and $\int_{0}^{1}f\left(  r\right)  ^{2}%
r^{d-1}dr=1/\omega_{d}$. The function $f_{\kappa,\sigma,\delta}$,
non-negative, is non-increasing; let us prove this by contradiction. Indeed,
if there are $r_{1}<r_{2}$ with $f_{\kappa,\sigma,\delta}\left(  r_{1}\right)
<f_{\kappa,\sigma,\delta}\left(  r_{2}\right)  $, by continuity of
$f_{\kappa,\sigma,\delta}$ (it is of class $W^{1,2}\left(  0,1\right)  $)
there exists a point $r_{\min}<r_{2}$ of minimum in $\left[  0,r_{2}\right]
$, with $f_{\kappa,\sigma,\delta}\left(  r_{\min}\right)  <f_{\kappa
,\sigma,\delta}\left(  r_{2}\right)  $. Given $l\in\left(  f_{\kappa
,\sigma,\delta}\left(  r_{\min}\right)  ,f_{\kappa,\sigma,\delta}\left(
r_{2}\right)  \right)  $, let $r_{l}^{+}$ be the minimum of all points $r>$
$r_{\min}$ such that $f_{\kappa,\sigma,\delta}\left(  r\right)  =l$; it exists
again by continuity of $f_{\kappa,\sigma,\delta}$. If $r_{\min}=0$, we
complete the contradiction as follows: for any such $l$ we introduce the
function $\widetilde{f}_{l}$ equal to $f_{\kappa,\sigma,\delta}$ on $\big[
r_{l}^{+},1\big]  $ and constantly equal to $f_{\kappa,\sigma,\delta} \big(
r_{l}^{+} \big)=l  $ in $\big[  0,r_{l}^{+} \big]  $. It is of class
$W^{1,2}\left(  0,1\right)  $, $\widetilde{f}_{l}\left(  1\right)  =0$,
$J\big(  \widetilde{f}_{l} \big)  <J\left(  f_{\kappa,\sigma,\delta}\right)
$. In itself this is not a contradiction yet because $a_{l}^{2}:=\int_{0}%
^{1}\widetilde{f}_{l}\left(  r\right)  ^{2}r^{d-1}dr$ is not equal to
$1/\omega_{d}$; but $a_{l}^{2}>1/\omega_{d}$, because $\widetilde{f}%
_{l}\left(  r\right)  =f_{\kappa,\sigma,\delta}\big(  r_{l}^{+}\big)
>f_{\kappa,\sigma,\delta}\left(  r\right)  $ in $\big[  0,r_{l}%
^{+}\big)  $; hence the function $f_{l}=\widetilde{f}_{l}/(|a_{l}| \sqrt{\omega_d} ) $
satisfies all the constraints and has the property
$J\left(  f_{l}\right)  <J\big(  \widetilde{f}_{l}\big)  $, hence $J\left(
f_{l}\right)  <J\left(  f_{\kappa,\sigma,\delta}\right)  $. If $r_{\min}>0$,
it is sufficient to introduce the maximum $r_{l}^{-}$ of all points
$r<r_{\min}$ such that $f_{\kappa,\sigma,\delta}\left(  r\right)  =l$ and
repeat the previous argument on $\big[  r_{l}^{-},r_{l}^{+}\big]  $ instead
of $\big[  0,r_{l}^{+}\big]  $.

\subsubsection{Proof of Theorem \ref{Thm eigen}}

Therefore $\lambda_{\kappa,\sigma,\delta}=\omega_{d}J\left(  f_{\kappa
,\sigma,\delta}\right)  $, where we know that $f:=f_{\kappa,\sigma,\delta}$ is
of class $W^{1,2}\left(  0,1\right)  $, non-negative and non-increasing,
$f\left(  1\right)  =0$, $\int_{0}^{1}f\left(  r\right)  ^{2}r^{d-1}%
dr=1/\omega_{d}$. We prove now several inequalities, some of them inspired by
the Poincar\'{e} inequality. First,
\begin{equation}\label{first-estim}
\aligned
\delta\int_{0}^{1}f^{\prime}\left(  r\right)  ^{2}r^{d-1}dr &  \geq\delta
\int_{1-\delta}^{1}f^{\prime}\left(  r\right)  ^{2}r^{d-1}dr\geq \left(
1-\delta\right)^{d-1} \bigg(  \int_{1-\delta}^{1}\left\vert f^{\prime}\left(
r\right)  \right\vert dr \bigg)^{2}\\
&  =\left(  1-\delta\right)^{d-1} \bigg(  -\int_{1-\delta}^{1}f^{\prime
}\left(  r\right)  dr\bigg)^{2} =\left(  1-\delta\right)  ^{d-1}f^{2}\left(
1-\delta\right)  .
\endaligned
\end{equation}
Second, since $f\left(  r\right)  =f\left(  1-\delta\right)  -\int%
_{r}^{1-\delta}f^{\prime}\left(  s\right)  ds$, we have%
\[
f\left(  r\right)  ^{2}\leq\left(  1+\gamma\right)  f^{2}\left(
1-\delta\right)  +\big(  1+\gamma^{-1}\big)  \int_{r}^{1-\delta
}f^{\prime}\left(  s\right)  ^{2}ds
\]
for all $\gamma>0$, hence we get, for $g\left(  \delta\right)  :=\int%
_{1-\delta}^{1}f\left(  r\right)  ^{2}r^{d-1}dr$,
\begin{align*}
\frac1{\omega_{d}} -g\left(  \delta\right)   &  =\int_{0}^{1-\delta}f\left(
r\right)  ^{2}r^{d-1}dr\\
&  \leq\frac{1+\gamma}{d}f^{2}\left(  1-\delta\right)  +\big(  1+\gamma
^{-1} \big)  \int_{0}^{1-\delta}\int_{r}^{1-\delta}f^{\prime}\left(
s\right)  ^{2}r^{d-1}dsdr.
\end{align*}
The double integral can be manipulated and shown to be equal to $\frac{1}%
{d}\int_{0}^{1-\delta}s^{d}f^{\prime}\left(  s\right)  ^{2}ds$, and a factor
$s$ in this integral can be bounded above by $1$. Notice also that, by
monotonicity of $f$, $g\left(  \delta\right)  \leq\delta f^{2}\left(
1-\delta\right)  $. We deduce%
\[
\int_{0}^{1-\delta}s^{d-1}f^{\prime}\left(  s\right)  ^{2}ds\geq\frac
{d}{\left(  1+\gamma^{-1}\right)  \omega_{d}}-\frac{1+\gamma}{1+\gamma^{-1}%
}f^{2}\left(  1-\delta\right)  -\frac{\delta d}{1+\gamma^{-1}}f^{2}\left(
1-\delta\right)  .
\]
Therefore, combining this inequality with \eqref{first-estim} and \eqref{J-functional} yields
\[
\lambda_{\kappa,\sigma,\delta}\geq\omega_{d}\left(  \frac{\kappa}{\delta
}\left(  1-\delta\right)  ^{d-1}-\sigma^{2}\frac{1+\gamma+\delta d}%
{1+\gamma^{-1}}\right)  f^{2}\left(  1-\delta\right)  +\frac{d}{1+\gamma^{-1}%
}\sigma^{2}.
\]
We now choose $\gamma$ such that
\[
\frac{1+\gamma+\delta d}{1+\gamma^{-1}}=\frac{\kappa}{\delta\sigma^{2}}\left(
1-\delta\right)  ^{d-1}%
\]
which is easily seen to be always possible. With this choice we have
$\lambda_{\kappa,\sigma,\delta}\geq\frac{d}{1+\gamma^{-1}}\sigma^{2}$. The
algebraic computations to complete the proof of the theorem are now elementary
but cumbersome, so let us give them only asymptotically as $\delta
\rightarrow0$. We thus have $\frac{1+\gamma}{1+\gamma^{-1}}=\frac{\kappa
}{\delta\sigma^{2}}$ which gives $\gamma=\frac{\kappa}{\delta\sigma^{2}}$,
hence $\lambda_{\kappa,\sigma,\delta}\geq\frac{\kappa d}{\kappa+\delta
\sigma^{2}}\sigma^{2}$, as stated in the theorem. It can also be rewritten as%
\[
\lambda_{\kappa,\sigma,\delta}\geq\frac{\frac{\kappa}{\delta}d}{\frac{\kappa
}{\delta}+\sigma^{2}}\sigma^{2}%
\]
which easily proves it is larger than $\frac{d}{2} \min\big(  \sigma^{2}%
,\frac{\kappa}{\delta} \big)  $ (if $\sigma^{2}\leq\frac{\kappa}{\delta}$,
then $\frac{\frac{\kappa}{\delta}d}{\frac{\kappa}{\delta}+\sigma^{2}}\geq
\frac{\frac{\kappa}{\delta}d}{2\frac{\kappa}{\delta}}=\frac{d}{2}$; similarly
in the opposite case).

\bigskip

\noindent \textbf{Acknowledgements:} The last named author would like to thank the financial supports of the National Key R\&D Program of China (No. 2020YFA0712700) and the National Natural Science Foundation of China (No. 11688101).


\end{document}